\documentstyle[prl,aps]{revtex}
\tighten
\begin{document}
\title{Stress-Energy Tensor for the Massless Spin 1/2 Field in Static Black Hole Spacetimes}
\author{Eric D. Carlson\thanks{Electronic address: {\tt ecarlson@wfu.edu}}, William H. Hirsch\thanks{Electronic address: {\tt hirswh1@wfu.edu}}, Benedikt Obermayer\thanks {Electronic address: {\tt b-obermayer@gmx.net}}, Paul R. Anderson\thanks{Electronic address: {\tt anderson@wfu.edu}},  Peter B. Groves\thanks{Electronic address: {\tt grovpb4@wfu.edu}}}
\address{Department of Physics, Wake Forest University, Winston-Salem, North Carolina, 27109}
\maketitle
\begin{abstract}
The stress-energy tensor for the massless spin 1/2 field is numerically computed outside and on the event horizons of both charged and uncharged static non-rotating black holes, corresponding to the Schwarzschild, Reissner-Nordstr\"om, and extreme Reissner-Nordstr\"om solutions of Einstein's equations.  The field is assumed to be in a thermal state at the black hole temperature.  Comparison is made between the numerical results and previous analytic approximations for the stress-energy tensor in these spacetimes.  For the Schwarzschild (charge zero) solution, it is shown that the stress-energy differs even in sign from the analytic approximation.  For the Reissner-Nordstr\"om and extreme Reissner-Nordstr\"om solutions, divergences predicted by the analytic approximations are shown not to exist.
\end{abstract}

Classically, the most general static black hole solution to Einstein's equations is spherically symmetric and is described by just two parameters, the mass ($M$) and charge ($Q$).  Quantum effects may be taken into account by solving the semi-classical backreaction equations~\cite{mtw},
\begin{equation}\label{semiclassical}
G_{\mu\nu} = 8\pi \langle T_{\mu\nu} \rangle \; ,
\end{equation}
where $\langle T_{\mu\nu} \rangle$ is the expectation value of the stress-energy tensor operator, which we shall hereafter refer to as the stress-energy tensor for the quantized fields.  It includes the effects both of physical particles and vacuum polarization.  A non-zero stress-energy tensor will alter the black hole's geometry.
% and hence its temperature and entropy~\cite{york,hky,ahwy}.
Even if Eq.~(\ref{semiclassical}) is not explicitly solved, finding the stress-energy tensor in a given background spacetime allows one to estimate the size of quantum effects in that spacetime.  For example, if the stress-energy diverges at the event horizon, quantum effects may prevent the existence of the black hole.  Such a divergence is predicted in one component of the stress-energy tensor by analytic approximations for massless spin 0, 1/2, and 1 fields in Reissner-Nordstr\"{o}m ($0<|Q|<M$) and extreme Reissner-Nordstr\"{o}m ($|Q|=M$) spacetimes~\cite{fz,zannias,huang,ahs} and, in one case, by a two dimensional calculation of the full renormalized stress-energy tensor for a massless spin 0 field~\cite{trivedi}.

Numerical computations of the stress-energy tensor for quantized fields in black hole spacetimes provide important tests of both analytic approximations and two dimensional calculations.  Previous numerical computations of the stress-energy of quantized fields in black hole spacetimes have primarily been for scalar fields in Schwarzschild ($Q=0$), Reissner-Nordstr\"{o}m, and extreme Reissner-Nordstr\"{o}m spacetimes~\cite{candelas,fawcett,hc,howard,jmo,ahs,ahl}.  While computations relating to scalar fields are important and provide some insight into the effects of quantized fields, it is important to also make computations with more realistic fields of nonzero spin.  Significant differences can exist between fields of zero versus nonzero spin~\cite{cht,tch}.  Because of difficulties in working with fields of higher spin in curved space, the only previous numerical calculation of the full renormalized stress-energy tensor in a black hole spacetime for a field of nonzero spin was for the massless spin $1$ field in Schwarzschild spacetime~\cite{jo}.    

In this paper we present the results of numerical computations of the stress-energy tensor for the massless spin 1/2 field in Schwarzschild, Reissner-Nordstr\"{o}m, and extreme Reissner-Nordstr\"{o}m spacetimes in the Hartle-Hawking-Israel state~\cite{hh,israel}, a thermal state at the black hole temperature.  Our results address the validity of the analytic approximations that have been derived for these spacetimes~\cite{bop,fz} as well as the question of whether the stress-energy is finite on the horizon.  We find that in Schwarzschild spacetime the analytic approximation for the field is very poor, not even giving the correct sign for the energy density on the horizon.  This is in contrast to the cases of massless scalar fields~\cite{hc,howard,ahs} and the spin $1$ field in Schwarzschild spacetime~\cite{jo} where it was found that analytic approximations were good both on and everywhere outside of the event horizon.  We also find that for Reissner-Nordstr\"{o}m and extreme Reissner-Nordstr\"{o}m spacetimes all components of the stress-energy tensor are finite on the event horizon.  This is in contrast to the analytic approximation of Frolov and Zelnikov~\cite{fz} which predicts a divergence in one component on the horizon.  It is also in contrast to the result found by Trivedi~\cite{trivedi} that one component of the stress-energy tensor for the conformally invariant field in two dimensions is divergent on the horizon of an extreme Reissner-Nordstr\"{o}m black hole.  However, it is in agreement with the numerical results found previously for massless scalar fields with arbitrary curvature couplings~\cite{ahs,ahl}.

In Ref.~\cite{gac} a method was given to compute the stress-energy tensor for the massless spin 1/2 field in
a general static spherically symmetric spacetime.  The method involves writing the stress-energy tensor in terms
of the Euclidean Green's function for the field and renormalizing using the method of point splitting~\cite{dewitt,christensen}.  An explicit expression for the Euclidean Green's function in a static spherically symmetric spacetime was derived in terms of mode solutions to the Euclidean Dirac equation.  The time dependent part of the modes and the angular dependent part were both given in terms of known functions.  The radial part was given in terms of solutions to a coupled set of ordinary differential equations.  For most spacetimes, including black hole spacetimes, these equations must be solved numerically.  The final expression for the stress-energy tensor consists of two parts both of
which are finite and separately conserved.  For one part the mode sums have been computed analytically for a general static spherically symmetric spacetime.  It has a trace equal to the trace anomaly and thus can serve as an analytic 
approximation for the entire stress-energy tensor.  It is equivalent to previous analytic approximations for the massless spin 1/2 field~\cite{bop,fz}.  The other part must usually be computed numerically.  

To simplify the numerical computation of the stress-energy tensor it is helpful to use the WKB approximation
for the high frequency and/or large angular momentum modes~\cite{hc,howard,ahs}.
The higher the order of the WKB approximation used the fewer modes that must be numerically computed to give an answer to a given accuracy.  However the number of terms in the WKB approximation grows rapidly with order so it is not practical to use too high an order.

The metric for a Reissner-Nordstr\"{o}m spacetime can be written 
\begin{equation}
ds^2 = -f(r) dt^2 + f(r)^{-1} dr^2 + r^2 d\Omega^2 \;,
\end{equation}
where
\begin{equation}
f(r)=1-{2M \over r} + {Q^2 \over r^2} \;.
\end{equation}
The temperature of the black hole is
\begin{equation}
2\pi T = {\sqrt{M^2-Q^2} \over 2 M^2 - Q^2 + 2 M \sqrt{M^2-Q^2}} \;.
\end{equation}
Recall that in the Hartle-Hawking-Israel state, the field is assumed to be in a thermal state at this temperature.  Note that the extreme Reissner-Nordstr\"om black hole is at zero temperature.

In Figs. 1 and 2 our results for Schwarzschild spacetime are displayed.  
As in the case of conformally invariant spin $0$~\cite{hc,howard} and $1$~\cite{jo} fields, the weak energy 
condition is violated near the horizon\footnote{With our conventions the energy density is $\rho = -{T_t}^t$.}.  
The analytic approximation is very poor near the horizon (although
it is accurate far from it), and  even gets the sign of the energy density wrong.  This is in contrast to the spin $0$ and $1$ cases where the analytic
approximation is good near the horizon~\cite{hc,howard,jo}.
 
As a check on this surprising result, we also computed the energy density at the horizon using radial point splitting with one point fixed on the horizon.  This method was developed by Candelas~\cite{candelas} who used it to compute the stress-energy tensor at the horizon for the conformally invariant scalar field in Schwarzschild spacetime.  It has the advantage that modes of only the two smallest allowed frequencies contribute to the mode sum.  The method in ~\cite{gac} involves splitting the points in the time but not the space direction.  The resulting expression for the stress-energy tensor contains a sum over modes with arbitrarily large frequencies.  Thus the two methods, while not completely independent, are significantly different.  The renormalization counterterms are also different for these two methods.  The energy density of the spin 1/2 field on the horizon in the radial point splitting calculation was found to agree to better than four digits with that obtained in the time point splitting calculation.

Some of our results for Reissner-Nordstr\"{o}m spacetimes are displayed in Fig.~3.  For non-zero charge $Q$, there is a well-known ambiguity in the analytic portion of the stress-energy tensor caused by renormalization~\cite{christensen,ahs,gac}.  This ambiguity does not arise for the Schwarzschild metric, nor does it exist on the horizon for $\langle {T^t}_t \rangle$, $\langle {T^r}_r \rangle$, or $\langle {T^\theta}_\theta \rangle$ in Reissner-Nordstr\"om or extreme Reissner-Nordstr\"om spacetimes.  For definiteness, we have chosen the renormalization parameter $\mu$ so that $\mu M=1$.  From Fig.~3, it is clear that the energy density on the horizon is negative and its magnitude increases with increasing $|Q|/M$.  Thus the weak energy condition is violated in these cases as well.
% Although not shown due to lack of space, the analytic approximation for $|Q| < M$ is also poor in these cases near the horizon.  It continues to be a good approximation far from the black hole.  For $|Q|=M$, the analytic approximations for $\langle {T^t}_t \rangle$, $\langle {T^r}_r \rangle$, and $\langle {T^\theta}_\theta \rangle$ matches on the horizon but is poor nearby.

One of the predictions of the analytic approximation for the massless spin 1/2 field is that one component of the stress-energy diverges on the event horizon for all Reissner-Nordstr\"om black holes.  The component is related to
the energy density seen by a freely falling observer passing through the event horizon and is explicitly 
\begin{equation}
{\langle {T^r}_r \rangle - \langle {T^t}_t \rangle \over f(r)}  \;.
\label{eq:rrmtt}
\end{equation}
For $|Q| < M$ the divergence in this component of the analytic approximation is logarithmic and of the form
\begin{equation}
   -\frac{Q^2}{20 \pi^2 r^6} \, \log\left[\mu^2 M^2 f(r) \right] \;,
\end{equation}
where $\mu$ is the renormalization parameter discussed above.  This divergence also occurs in the case $|Q|=M$.  However in that case there is a term in the analytic approximation for the component (\ref{eq:rrmtt}) with a stronger divergence.  It is
\begin{equation}
  \frac{M^2(32 M - 23 r)}{180 \pi^2 r^6 (r - M)} \;.
\end{equation}
This is also the type of divergence found by Trivedi~\cite{trivedi} for the same component of the full renormalized stress-energy tensor of a conformally invariant scalar field in a two dimensional extreme Reissner-Nordstr\"om spacetime.

However, when we numerically computed the component (\ref{eq:rrmtt}) for the massless spin 1/2 field in Reissner-Nordstr\"om and extreme Reissner-Nordstr\"{o}m spacetimes, we found no evidence that this component of the stress-energy tensor diverges on the event horizon.  This is in agreement with results found for massless scalar fields with arbitrary curvature couplings in these spacetimes~\cite{ahs,ahl}.  Some of our results are shown in Fig.~4.

In summary the full renormalized stress-energy tensor for the massless spin 1/2 field has been numerically computed 
in Schwarzschild, Reissner-Nordstr\"om, and extreme Reissner-Nordstr\"om spacetimes in the case that the field is in the Hartle-Hawking-Israel state.  The analytic approximation for the stress-energy tensor in Schwarzschild spacetime is found to be very poor near the event horizon.  Divergences in the stress-energy, predicted by the analytic approximations for Reissner-Nordstr\"om and extreme Reissner-Nordstr\"om, are found not to occur.  Thus it is clear that the analytic approximations for the massless spin 1/2 field in static black hole spacetimes are of extremely limited use and must usually be supplemented by numerical calculations of the full renormalized stress-energy tensor.

\acknowledgements

This work has been supported in part by grant number PHY-0070981 from the National Science Foundation.

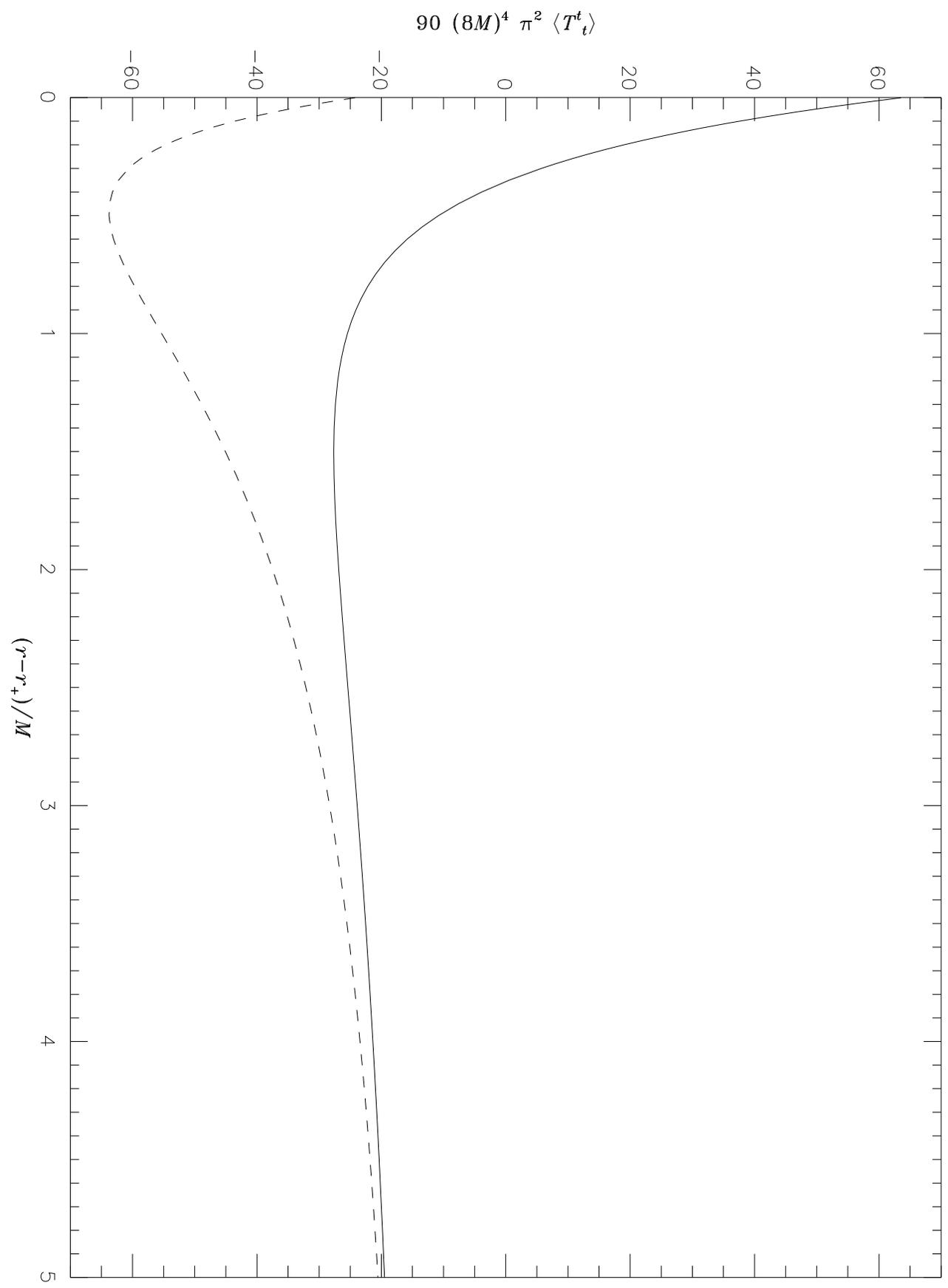
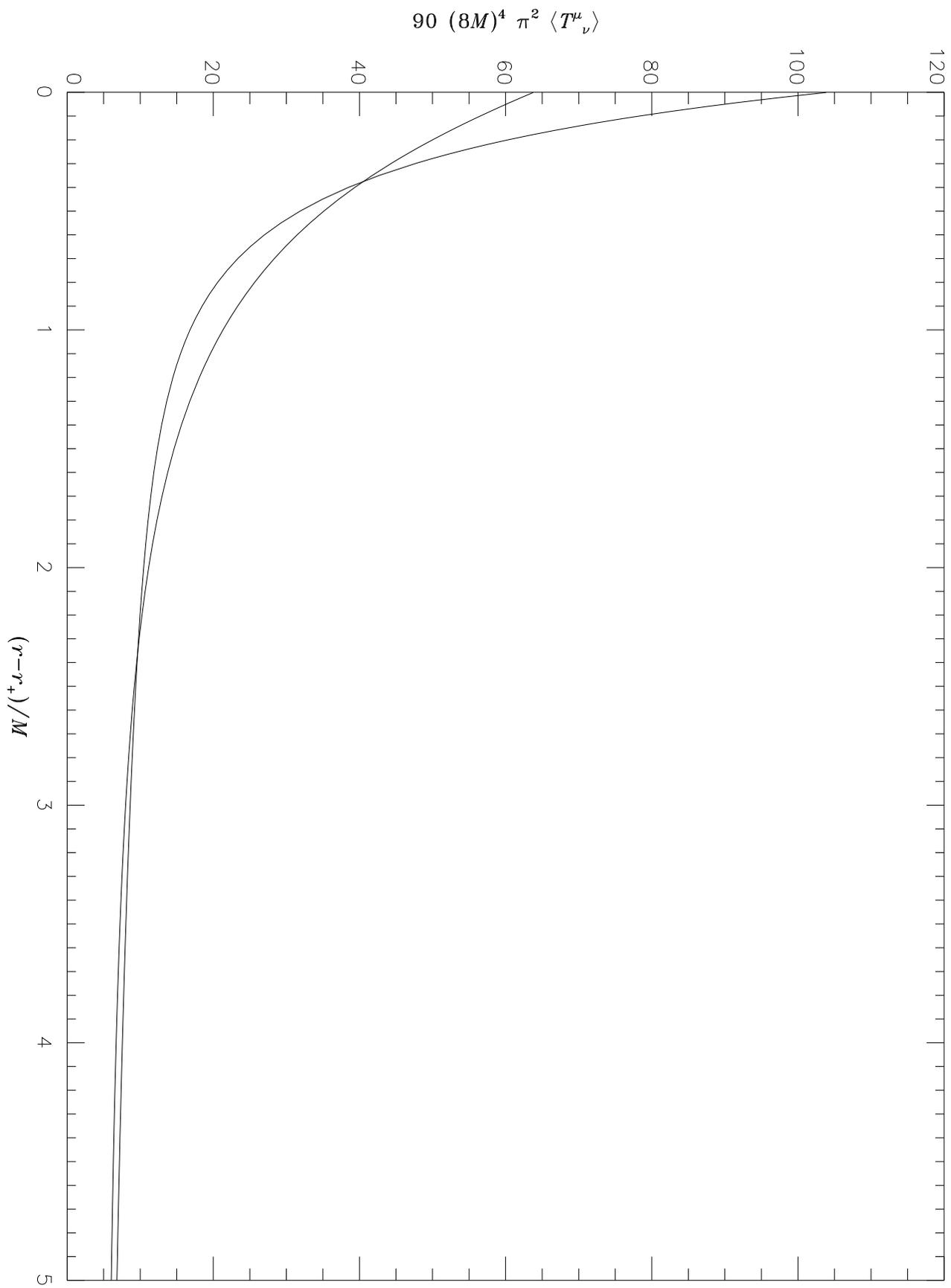
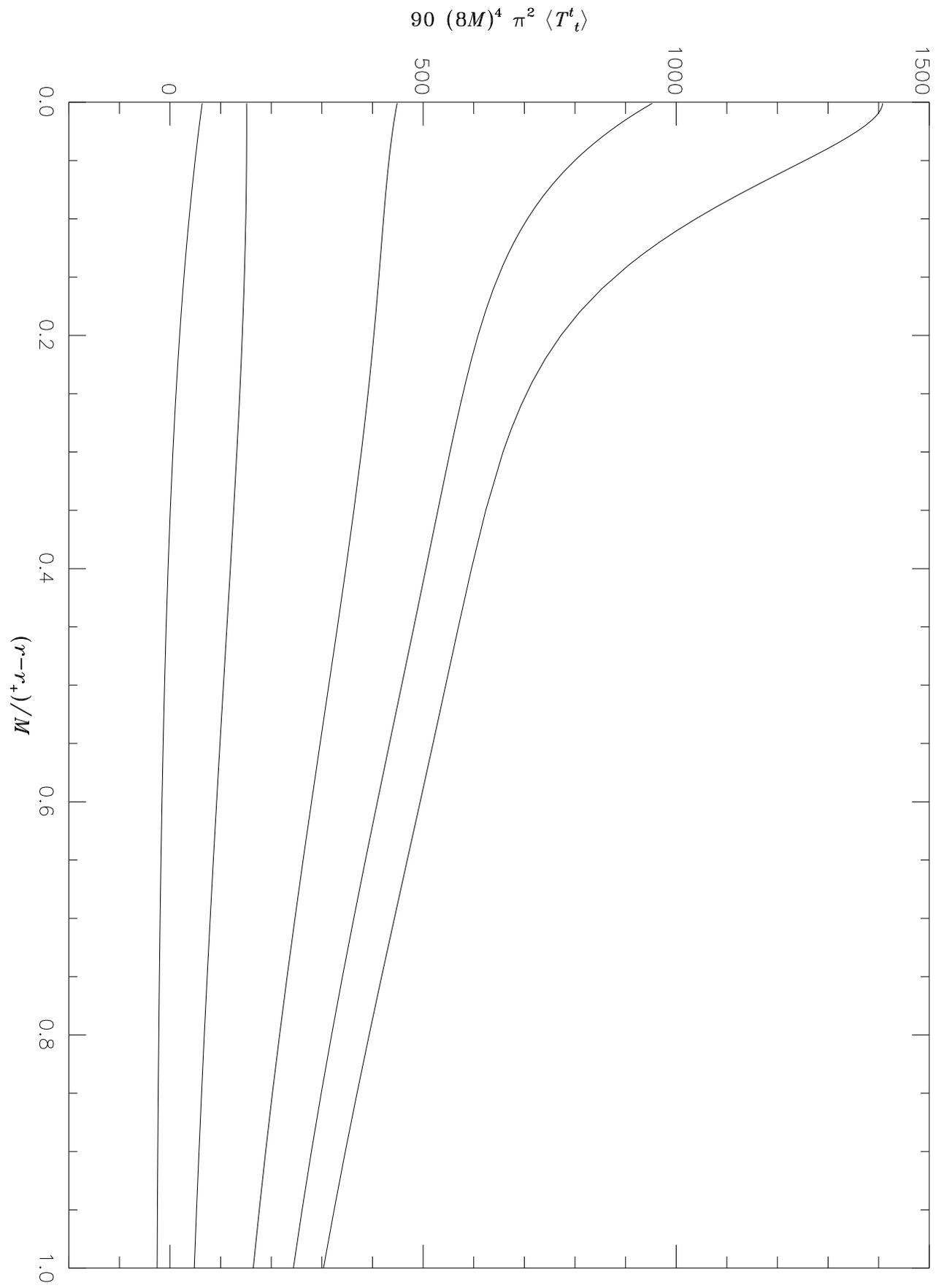
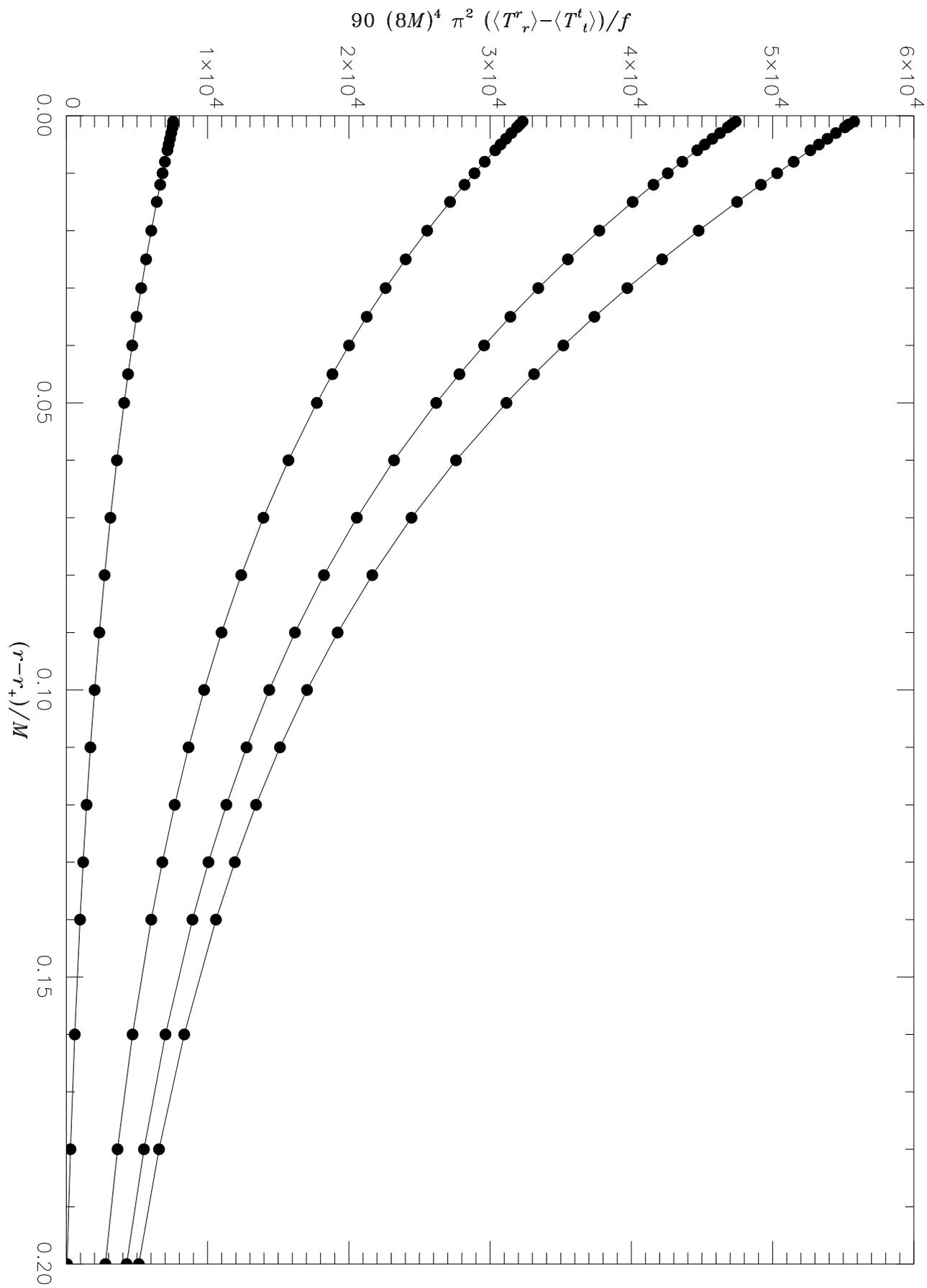
\end{document}